\documentclass[11pt]{article}
\usepackage{amsmath}
\usepackage{amsfonts}
\usepackage{amssymb}
\usepackage{graphicx}
\usepackage{subfigure}

\def\be\begin{equation}
 \def\ee{\end{equation}}
\def\bea{\begin{eqnarray}}
\def\eea{\end{eqnarray}}

\def\l{\label}
\begin{document}
\begin{center}
\LARGE {Reconstructing f(R) model from Holographic DE:   Using the observational evidence }
\end{center}
\begin{center}
{\bf $^{a,b} $ Kh. Saaidi\footnote{ksaaidi@uok.ac.ir}},
{\bf $^{c}$A.Aghamohammadi\footnote{a.aghamohamadi@iausdj.ac.ir}},\\
{\it$^a$Department of Physics, Faculty of Science, University of
Kurdistan,  Sanandaj, Iran}\\
{ \it $^b$Department of Physics, Kansas State University,116 Cardwell Hall, Manhattan, KS 66506, USA.}\\
{\it $^c$Faculty of Science,  Islamic Azad University Sanandaj Branch, Sanandaj, Iran }
\end{center}
 \vskip 1cm
\begin{center}
{\bf{Abstract}}
\end{center}

We investigate the corresponding relation between  $f(R)$ gravity and an  interacting  holographic dark energy. By obtaining conditions needed for  some observational evidence such as, positive acceleration expansion of universe, crossing the phantom divide line and validity of thermodynamics second law in an interacting HDE model and corresponding it with $f(R)$ mode of gravity  we find a viable $f(R)$ model which can explain the present universe. We  also obtain the explicit evolutionary
forms of the corresponding scalar field, potential and scale factor of universe.
 \\

\newpage
\section{Introductions}

 Observational data\cite{1,2,3, 4, 5}, indicates  that the current expansion of universe is accelerating. Several attempts have been made to justified the current accelerated expansion of the universe \cite{14, 15,16,18,19}.  One is the presentation of an unknown energy form  which is called dark energy.  An alternative approach is the modification of the  gravitational theory e.g. $f(R)$ gravity in which $f(R)$ is  an arbitrary function of the scalar curvature R \cite{14,17, n10}. Recent various observational data imply that  the density of  matter (ordinary matter $+$ dark matter$+$ radiation), $\Omega_m=0.27$ and the density of dark energy, $\Omega_{\Lambda}=0.73$  have capable value today (coincidence problem), beside based on recent data, the equation of state parameter may evolve from $\omega>-1$ (non-phantom phase) in the past to $\omega<-1$ (phantom one) at the present epoch. One way to explain these data, is to consider dynamical dark energy with proper interaction  with matter\cite{la}.\\
In the quantum field theory $\rho_{\Lambda}$ is regarded as zero-point energy density and defined  based on $L$, the  size of the current universe, (dubbed  the   holographic dark energy ) as follow
 \begin{equation} \label{1}
\rho_{\Lambda}= 3c^2 M_p^2L^{-2},
\end{equation}
where $c^2$ is a numerical constant of order unity and $M_p=1/\sqrt{8\pi G}$ is  the reduced Planck mass where $G$ is the Newtonian gravitational constant. Different choices may be adopted for the infrared cutoff of the universe, e.g., Hubble horizon, particle horizon, future event horizon \cite{qg}. In a non interaction model, if we take the particle horizon as the infrared cutoff, the accelerated expansion of universe cannot be explained\cite{2c}, and if the Hubble horizon does chooses as the cutoff, then an appropriate equation of state parameter for dark matter cannot be derived \cite{sd}. By taking, the future event horizon as the cutoff, the present expansion of universe may be explained but the coincidence problem still remains unsolved. This problem   may be alleviated by considering suitable interaction  between  dark matter    and holigraphic dark energy.\\
In this paper,  we  consider a flat friedmann-Robertson- Walker universe  and assume that the universe is composed of two interacting perfect fluids, the holographic dark energy and the  matter.  We assume  the infrared cutoff  to be  a combination of the future and particle event  horizon. After some general debate about the properties of the model, we discuss the required conditions needed to cross the phantom divide line in the $f(R)$ model. We show that this crossing imposes some relations between the parameters of the model.\\
In this paper  we well review $f(R)$  model of gravity  and  make a correspondence  between $f(R)$ model and an interacting  holographic dark  energy model. By investigating the conditions which are needed for describing the present universe, we can obtain a viable $f(R)$ model of gravity.

 \section{Description and general  properties of  the model}

The equation of motion for  the $f(R)$  model is
\begin{equation}\label{2}
R_{\mu\nu}f '-\frac{1}{2}fg_{\mu\nu}+\left( g_{\mu\nu}\Box-\nabla_{\mu}\nabla_{\nu}\right)f '=8\pi GT_{\mu\nu},
\end{equation}
where a  prime represents  the derivative with respect to  the curvature scalar $R$ and $\Box$ is the covariant  D'Alembert operator ($\Box\equiv \nabla_{\alpha}\nabla^{\alpha}$). We  will assume  dark energy and  cold dark matter  perfect fluids with stress-energy   tensor given by
\begin{eqnarray}\label{3}
T_{\mu\nu}=-g_{\mu\nu}p+\left(\rho+p \right)u_{\mu}u_{\nu},
\end{eqnarray}
 where $\rho$ and $p$  are  the energy density  and  pressure of the fluid and $u^{\mu}=\left( 1, 0, 0, 0\right) $ is its normalized four-velocity in  co-moving  coordinates. The dark energy  component  has  pressure $p_d$ and energy  density $\rho_d$  and the cold dark matter  component has zero pressure  and energy  density  $\rho_m$. The  stress-energy   tensor is covariantly conserved. \\

  The trace of     equation  (\ref{2}) gives an  equation of motion for the new scalar degree of freedom (compared to Einsteinian general relativity),  \cite{9n, ka},
\begin{eqnarray}\label{4}
3\Box f '={8\pi GT}+  2f-Rf'  ,
\end{eqnarray}
  where $T$ is the trace of the stress-energy tensor. It is helpful to redefine  the scalar degree of freedom through
     \begin{eqnarray}\label{5}
\phi=f '-1.
\end{eqnarray}
Then Eq. (\ref{4}) can be reexpressed as an   equation of motion for a canonical dimensionless scalar field $\phi$ with a force term ${\cal F}$ and potential $V$,
\begin{eqnarray}\label{6}
\Box \phi&=& V'(\phi)-{\cal F},\\
3V'(\phi)&=&  2f-Rf '\l{7},
\end{eqnarray}
where the force term that drives the scalar field $\phi$ is proportional to the  trace of the stress-energy tensor, ${\cal F}=-8\pi GT/3$.\\

Now we consider   a homogeneous and spatially-flat  spacetime  with FLRW line element
   \begin{equation}\label{8}
ds^2= dt^2-a^2(t)(dx^2 +dy^2+dz^2),
\end{equation}
where $a(t)$ is the scale factor. The $tt$ component of the  gravitational equations (\ref{2}), for the metric (\ref{8}), can be simplified to
\begin{eqnarray}\label{9}
H^2+H{d \over dt}\left(\ln f' \right)-\frac{1}{6}\left( \frac{f-R f'}{f'}\right)=\frac{8\pi G}{3f'}\rho_m.
\end{eqnarray}
The Friedmann  equation, (\ref{6}), can be  written  in a  somewhat more conventional form as
   \begin{eqnarray}\label{10}
H^2=\frac{8\pi G }{3\left(1+\phi \right) }\left( \rho_m+\rho_d\right),
\end{eqnarray}
where we   assume that  the new scalar degree of freedom  behaves like dark energy with  dark energy density
\begin{equation}\label{11}
\rho_d= -\frac{3\left(1+\phi \right)  }{8\pi G}\left[H { d(\ln f')\over dt}-\frac{1}{6}\left( \frac{f-Rf '}{f '}\right)\right].
\end{equation}
Also, one can write the Friedmann equation as
\begin{eqnarray}\label{12}
\Omega_m+\Omega_d=1,
\end{eqnarray}
where the density parameters $\Omega_m={\rho_{m}}/{\rho_c}$, $\Omega_d={\rho_{d}}/{\rho_c}$,  and the critical energy density is
\begin{equation}\l{13}
  \rho_c={3H^2(1+\phi) \over 8\pi G}.
  \end{equation}

   From the  Friedmann equation  (\ref{10}) and conservation of  stress-energy tensor,  we have
 \begin{eqnarray}\label{14}
\dot{H}=-\frac{4\pi G\left( \rho_t+p_d\right)}{1+\phi} -\frac{\dot{\phi }H}{2(1+\phi)},
\end{eqnarray}
where   $\rho_t = \rho_m + \rho_d$.
The vanishing of  the covariant  divergence of the stress-energy  tensor for the whole system gives the conservation equation in the metric (\ref{8}),
 \begin{eqnarray}\label{15}
\dot{\rho}_t+3H\left( \rho_t+p_d\right)=0.
\end{eqnarray}
But, because of interactions between the two components, each individual  component is not necessarily  conserved. So, one can write
\begin{eqnarray}\label{16}
\dot{\rho_d} +3H\left( \rho_d+p_d\right)&=&-Q,\\
\dot{\rho_m} +3H\rho_m  &=& Q.\l{17}
\end{eqnarray}
We consider different forms of $Q$ below.
 A number of different models have been proposed for  dark energy. Here we want to investigate the holographic dark energy model and see if it can be related to the $f(R)$ model. Holographic  dark energy is described  in terms of an  infrared cut-off length, $L$, and the energy density  is defined as
   \begin{equation}\l{18}
   \rho_d = {3c^2 M_p^2 \over L^2}.
   \end{equation}
 where $c^2$ is a constant of order unity  and $M_p$ is the Plank mass.
  This is motivated by      quantum theory of
gravity considerations, in particular  the holographic principle \cite{2c, sd, ss}. It was shown in
\cite{ss1} that in quantum field theory the UV cutoff $\Lambda$ should be related to the IR cutoff $L$ due to a  limit
set by forming a black hole with Schwarzschild radius $L$. If $\rho_d =\Lambda^4$ is the vacuum energy density of the  UV cut-off scale, the
total energy in volume $L^3$ should not exceed the mass of the system-size black hole. This means that $L^3\rho_d \leq M_p^2 L$.
 So for  the largest cut-off $L$,   one can define the holographic dark energy  as (\ref{18}).
   From Eqs. (\ref{18})  and  (\ref{13}),   the  density parameter  of holographic dark energy can written as
 \begin{eqnarray}\label{19}
\Omega_d=\frac{c^2}{\left( 1+\phi \right)H^2 L^2 } .
\end{eqnarray}

The IR cut-off, $L$, is presumably determined by the available length scale. To retain generality, we assume that it is a linear  combination of  the particle horizon, $R_p$, and the future  event horizon, $R_f$, i.e., we choose $L$ to be
\begin{eqnarray}\label{20}
L=\alpha R_f+\beta R_p,
\end{eqnarray}
where
\begin{eqnarray}\label{21}
R_f=a(t)\int_t^{\infty}\frac{\mathrm{d}t}{a(t)}, \qquad R_p = a(t)\int_{t_{\rm min}}^{t}\frac{\mathrm{d}t}{a(t)},
\end{eqnarray}
here $t_{\rm min}$ is the time when the particle was created, and $0\leqslant \alpha, \beta \leqslant 1$ and $\alpha + \beta = 1$.
For $\alpha=1, \, \beta=0$ we get $L=R_f$ while $\alpha=0, \, \beta=1$  gives $L=R_p$.\\

Taking the      time derivative  of (\ref{18}),  and using  (\ref{16}),  one can obtain the equation of state parameter
$\omega_d = p_d/\rho_d$,
   \begin{equation}\l{22}
   \omega_d = -{1 \over 3}  {\biggr[}1+ {Q \over H\rho_d} -{2(\beta-\alpha) \over HL}{\biggl ]}.
   \end{equation}
    To progress,  we have to specify the interaction term $Q$.  A generic form of $Q$ is
not available.  Three forms which are often discussed in the literature are   $Q = 3b^2H\rho_d, 3b^2H\rho_m, 3b^2H\rho_t $, where $b^2 $ is a constant which  has to be positive, because following  the second law of thermodynamic, energy transfer  can only  be from dark energy  to cold dark mater. These three forms of interaction give  almost the same result, so for definiteness,  we  choose
    \begin{equation}\l{23}
    Q = 3b^2 H\rho_d .
    \end{equation}
     Using (\ref{19}), (\ref{22}) and (\ref{23}), we find
     \begin{equation}\l{24}
   \omega_d = -{1 \over 3} {\biggr[}1+ 3b^2 -{2(\beta-\alpha) \over c} \sqrt{(1+\phi)\Omega_d} {\biggl ]}.
   \end{equation}

As  mentioned in the  Introduction,   observational data indicate  that the current cosmological  expansion  is  accelerating. In the fluid model  this accelerated expansion  requires    $\omega_d < -{1 / 3 }$. This constraint  results in
\begin{equation}\l{25}
{2(\beta-\alpha) \over c} \sqrt{(1+\phi)\Omega_d} < 3b^2.
\end{equation}
Defining  a positive constant  $0 < k_0 <1 $, we can rewrite (\ref{25}) as
\begin{equation}\l{26}
{2(\beta-\alpha) \over c} \sqrt{(1+\phi)\Omega_d} = 3k_0b^2.
\end{equation}

   The second law of  thermodynamics requires that the entropy S increasing with time then,    $\dot{S} > 0$. We assume  $S$ is the  entropy  attribute to the surface area $A =4\pi L^2$, where $L$ is the infrared cut-off length   appearing  in (\ref{18}). Also,  making use of
 Nother charge method, one  can obtain the  entropy in the $f(R)$ model of  gravity  for a horizon with  surface $A=4\pi L^2$ as \cite{sad}
   \begin{eqnarray}\label{27}
S=\frac{A f '(R)}{4} = \pi L^2 (1 + \phi),
\end{eqnarray}
Then, considering the thermodynamics second law,  the time derivative of entropy,  $S$,  should be
\begin{equation}\l{28}
{\dot{S} \over \pi L^2 } = {\biggr [}2H + {\dot{\phi} \over 1+\phi} + {2(\beta -\alpha) \over L}{\biggl ]}\geq 0.
\end{equation}
We set (\ref{26}) as
\begin{equation}\l{29}
 {\biggr [}2H + {\dot{\phi} \over 1+\phi} + {2(\beta -\alpha) \over L}{\biggl ]}= s_0,
\end{equation}
where $ 0 \leq s_0 $.  $ s_0 =0 $ is when the accelerating expansion of the horizon of  universe be adiabatic. Here we assume $s_0 >0$ and then
 \begin{equation}\l{30}
  {2(\beta -\alpha) \over H L}= s_0-{\dot{\phi} \over H(1+\phi)} -2 ,
\end{equation}
By making use of (\ref{26}) and (\ref{30} ),  we have

\begin{equation}\l{31}
 {\dot{\phi} \over H(1+\phi)} = s_0-3k_0b^2 -2  = \theta_0,
\end{equation}

On the other hand, based on recent data, the dark energy component seems to have an equation of state parameter  $\omega_d < -1$ at the present epoch, while $\omega_d > -1$  in the past \cite{u12}. Therefore, we expect the equation of state parameter cross the phantom divide line, then  when $\omega =-1$,  the   crossing is allowed. So by implying the phantom crossing line constraint on $\omega$,  (\ref{24}), we have
 \begin{equation}\l{32}
  {2(\beta -\alpha) \over H L}= 3b^2 -2.
\end{equation}
 From (\ref{26}) and (\ref{32}) we have
 \begin{equation}\l{33}
 0 < k_0 = 1 -{2 \over 3b^2} < 1,
 \end{equation}
 this show that $0 < {2/ 3b^2} <1$. This means that one of the constant can be omit.  Moreover, to cross $\omega_d =-1$, $\dot{\omega}_d$ must be negative at the transition time,  $\omega_d =-1$.
 So by using  (\ref{32}) and  time derivative of (\ref{24}) we have
 \begin{equation}\l{34}
\dot{ \omega} = -{\biggr (}{\dot{H} \over H^2} + {3\over 2}b^2{\biggl )}{\biggr (} b^2 -{2\over 3}{\biggl )}<0.
 \end{equation}
 From (\ref{33}), we have $b^2 > {2 /3}$, so that the relation (\ref{34}) is satisfied when
 \begin{equation}\l{35}
 {\dot{H} \over H^2} > -{3 \over 2}b^2
 \end{equation}
So solving (\ref{35}), gives
\begin{equation}\l{36}
 H=  {h_0 \over 1 + h_0 \gamma t},
 \end{equation}
where $\gamma = {3 \xi_0b^2/2}$, $\xi_0$ is an arbitrary  constant which satisfy $0< \xi_0 <1$ condition. We assume $H_0t_0 \sim 1$ ($H_0$ and $t_0$ are Hubble parameter in the present time respectively) then $h_0 = {H_0 /(1-\gamma)}$.  By making use of (\ref{36}) and (\ref{31}) we can easily find the scale factor of universe and the scalar field $\phi$ as
\begin{eqnarray}\l{37}
a(t) &= & a_0 (1+h_0\gamma t)^{1 \over \gamma}, \\
\phi(t) &=& -1 + \phi_0 (1+h_0\gamma t)^{ \theta_0 \over \gamma},\l{38}
\end{eqnarray}
 where $a_0 = { (1-\gamma)^{1/ \gamma}}$( we assume that the scale factor in the present time is equal to 1, $a(t_0) =1$)  and $\phi_0$  are the integration constants. It is clearly seen that, from (\ref{37}), at the early time $a(0) \neq 0$ and  then we have bouncing in the beginning of the universe.
It is well known that     the Rcci scalar in flat FLRW is as
    \begin{equation}\l{39}
    R = 6{\bigr[}{\ddot{a} \over a } + ({\dot{a} \over a })^2{\bigl ]},
    \end{equation}
    then
    \begin{equation}\l{40}
    R =  {R_0 \over (1+h_0\gamma t)^2},
    \end{equation}
   where $R_0 = 6h^2_0(1-2\gamma) $. So that using (\ref{5}), (\ref{39}) and (\ref{40}), one can obtain a viable $f(R)$,  which allow the crossing from $\omega =-1$, as
    \begin{equation}\l{41}
    f(R) = f_0  +C R^{1-\epsilon},
    \end{equation}
    here $C_0 = {\phi_0 R_0^{\theta / 2\gamma}/ \epsilon}$, $\epsilon={\theta /2\gamma}$  and   $f_0$ is the constant of integration which can be as well as cosmological constant, $\Lambda$. By making use of chameleon mechanism this kind of $f(R)$ model has been studied in \cite{k1} and they show that this form of $f(R)$ model, is viable and satisfy the observational constraints  solar system. Also using (\ref{38}) and (\ref{41}) one can rewrite (\ref{7}) as
    \begin{equation}\l{42}
    V(\phi) = V_0 +V_1 \phi + {V_2 \over (1+\phi)^{2\gamma\theta}},
        \end{equation}

where $V_0$ is a constant of integration, $V_1 = 3f_0/2 $ and
$$V_2 = {\epsilon(1+\epsilon)\phi_0^{1\over \epsilon} R_0 \over (1-\epsilon)(2\epsilon -1)}$$

\section{Conclusion}
The HDE model is an attempt for probing
the nature of DE within the framework of quantum gravity \cite{k}. In this work  we used of  HDE model which is in interaction with DM in the flat FLRW universe.
 We established a correspondence between
the interacting HDE model with the $f(R)$ model of gravity  in the flat FLRW universe. These correspondences are important
to understand how  different models which  have been  candidated for explanting the present universe,  are mutually related to each other.  However, by taking account an infrared  cutoff as a combination of  particle and future event horizons and
 using the HDE energy density, we obtained the EoS parameter for the interacting HDE. Using equations derived  for equation of state parameter
of  holographic dark energy and its time derivative, condition required for crossing
the phantom divide line was derived. Also  the condition of validity
of thermodynamics second law for the infrared cutoff was obtained.   Thus we studied the evolving behavior of the interacting HDE and implying some observational evidence such as, positive acceleration expansion of the universe ($\omega <-1/3$ and $q<0$), crossing the phantom divide line ($\omega <-1$) and  validity os second law of thermodynamics for an interacting model of HDE,  we  reconstructed the $f(R)$ model which describe accelerated expansion of the universe. We  also obtained the explicit evolutionary
forms of the corresponding scalar fields, potential and scale factor of universe.
\section{Aknowledgement}
 The work of  Kh. Saaidi have been supported financially  by  University of Kurdistan, Sanandaj, Iran, and  he would like thank to the University of Kurdistan for supporting him in sabbatical period.

\end{document}